\documentclass[a4paper,11pt]{article}

\usepackage{jcappub} 
\usepackage{bm}
\usepackage{multirow} 
\usepackage{url}
\usepackage{color}
\usepackage[normalem]{ulem}
\usepackage[T1]{fontenc}
\usepackage[dvipsnames,table]{xcolor}
\usepackage{ragged2e}

\newcommand{\be}{\begin{equation}}
\newcommand{\ee}{\end{equation}}
\newcommand{\bea}{\begin{eqnarray}}
\newcommand{\eea}{\end{eqnarray}}
\newcommand{\gev}{\rm{GeV}}
\newcommand{\mev}{\rm{meV}}

\hypersetup{urlcolor=RedViolet,
	    citecolor=ForestGreen ,
	    linkcolor=RoyalBlue, 
	    colorlinks=true
}

\let\vec\mathbf

\title{Directional detection of meV dark photons with Dandelion}

\author[a]{C. Beaufort,}
\author[b,a]{M. Bastero-Gil,}
\author[a]{A. Catalano,}
\author[a]{D-S. Erfani-Harami,}
\author[a]{O. Guillaudin,}
\author[a]{D. Santos,}
\author[a]{S. Savorgnano,}
\author[a]{and F. Vezzu}

\affiliation[a]{Laboratoire de Physique Subatomique et de Cosmologie, Universit\'{e} Grenoble-Alpes, CNRS/IN2P3,\\38000 Grenoble, France}
\affiliation[b]{Departamento de F{\'i}sica Te{\'o}rica y del Cosmos, Universidad de Granada,\\Granada-18071, Spain}

\emailAdd{cyprien.beaufort@lpsc.in2p3.fr}
\emailAdd{mbg@ugr.es}
\emailAdd{daniel.santos@lpsc.in2p3.fr}

\abstract{This paper presents Dandelion, a new dish antenna experiment searching for dark photons (DPs) with masses around the meV that will start acquiring data by the end of 2023. A spherical mirror acts as a conversion surface between DPs and standard photons that converge to a matrix of $418$ Kinetic Inductance Detectors cooled down to $150~\rm{mK}$. A tilt of the mirror at $1~\rm{Hz}$ moves the expected signal over the pixels thus enabling a continuous background measurement. The expected signal has two modulations: a spatial modulation providing a directional signature for the unambiguous discovery of a DP, and an intensity modulation allowing the determination of the polarization of the DP. For masses near the meV, the inflationary production of longitudinal and transverse DPs are mutually excluded, thus the polarization determination by Dandelion could shed a new light on the inflation phase of the early universe. A first Dandelion prototype operating for 30 days would improve by more than one order of magnitude the current exclusion limits on DPs at the meV mass scale and would probe this region with an unprecedented discovery potential based on directional detection.}

\begin{document}
\maketitle
\flushbottom

\section{Introduction}

The direct detection of Dark Matter (DM) particles relies on complementary detection strategies to explore the "galaxy" of postulated DM candidates \cite{APPEC2021, PDG2020}. While the observation of a DM-like signal in a detector would represent an important forward step, the proper demonstration that such a signal is actually associated with local DM remains an additional challenge and is often left for the future. Directional detection overcomes this limitation by designing experiments able to measure unambiguous DM signatures distinguishable from any background. The directional strategy consists in correlating the measured signal to the detector's motion through the galactic DM halo \cite{Spergel1987, Mayet2016}.

The dark photon (DP) is a hypothetical massive spin-1 boson predicted in multiple extensions of the Standard Model (SM) and arising from a new $U(1)$ symmetry under which SM fields are uncharged \cite{Jaeckel2010, Jaeckel2012, Fabbrichesi2020}. There is a large variety of cosmological production mechanisms, thermal and non-thermal, for the DP as a suitable cold DM candidate \cite{Nelson2011, vectorDM, vectorDMown, vectorDM1, vectorDM2, vectorDM3}. Besides this compelling property, the interest in the dark photon is also enhanced as being an appropriate \textit{vector portal} to the dark sector \cite{Alexander2016, Battaglieri2017}. Since the DP is the gauge boson of an Abelian $U(1)$ symmetry, it interacts with the standard photon via a kinetic mixing \cite{Arias2012}:
\begin{equation}
   {\cal L }= -\frac{1}{4}F_{\mu \nu}F^{\mu \nu}-\frac{1}{4}X_{\mu \nu}X^{\mu \nu}-\frac{\chi}{2}F_{\mu \nu}X^{\mu \nu}+\frac{1}{2}m_X^2X_\mu X^\mu + e J_\mu^{em} A^\mu \,,
\end{equation}
where $A^\mu$ is the standard photon with field strength $F^{\mu \nu}$, $X^{\mu}$ is the DP with field strength $X^{\mu \nu}$, and $J_\mu^{em}$ is the electromagnetic current. The kinetic mixing, which is quantified by the dimensionless parameter $\chi$, opens the window for detection by the conversion of a DP into a standard photon. 

DPs are constrained experimentally by numerous experiments; we refer to \cite{Caputo2021} for a comprehensive review. The kinetic mixing parameter is however poorly constrained for a DP mass around $1~\rm{meV}$. We will see that a DP with a mass near $1~\rm{meV}$ would provide a clear signature for distinguishing between production mechanisms, hence serving as a probe for early universe physics. These two main motivations, in parallel with developments of new detectors, result in an increasing experimental interest for meV DPs. Multiple experiments with diversified strategies of detection have set limits on DPs with masses close to the meV \cite{Redondo2008, Ehret2010, Schwarz2015, Knirck2018, An2020, Fan2022} and efforts will continue to explore this region in the near future. In particular, the BREAD collaboration \cite{BREAD2021} aims to build a detector with significantly improved sensitivity compared to existing limits. 

In comparison with cited experiments, the main contribution of our detector lies in its directional feature for identification of the DP\footnote{Note that the BREAD collaboration considers directional detection as a future upgrade \cite{BREAD2021}.}. We rely on the concept of dish antenna for DP searches that was proposed in \cite{Horns2012} and refined in \cite{Jaeckel2013, Jaeckel2015, Jaeckel2017} for directional detection. A spherical metallic mirror acts as a conversion surface from DPs to ordinary photons by conservation of the parallel electric field at the surface. The energy of the emitted photons can be considered equal to the mass of the DPs. The DP-induced photons are emitted almost perpendicularly to the mirror surface and thus concentrate in a spot at a distance $R$ from the mirror, with $R$ the curvature radius, where one can place a photon detector. The emission deviates from the normal of the mirror by an angle $\psi = |\vec{p}_{||}|/m$ where $\vec{p}_{||}$ is the DP momentum component parallel to the surface of the mirror. This deviation results in a spatial modulation of the signal which is correlated to the detector's motion through the galactic DM halo, thus providing a directional signature for an unambiguous DP detection \cite{Jaeckel2013}. Assuming that all the local DM is made of DP, the DP-induced photon power at time $t$ and a position $(x,y)$ from the center of a photon detector placed at a distance $R$ from the mirror is given by \cite{Horns2012}:
\begin{equation}
P(x, y, t; E_\gamma) = \chi^2\,\rho_{\mathrm{CDM}}\,\eta(E_\gamma)\,A_{\mathrm{mirr.}}\,I\big(x, y,t)\,\cos^2\alpha(t)\,,
	\label{eq:power}
\end{equation}
with $\rho_{\mathrm{CDM}}$ the local DM density, $\eta(E_\gamma)$ the detection efficiency for a photon of energy $E_\gamma$, $A_{\mathrm{mirr.}}$ the surface of the mirror, $I\big(x, y,t)$ the normalized signal spread on the detector, and $\alpha(t)$ the angle between the DP polarization and the tangent of the mirror.

\begin{figure}[t]
    \centering
	\includegraphics[width=0.7\linewidth]{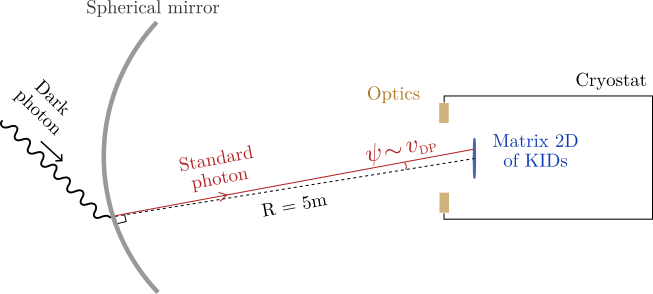}
	\caption{Schematics of the working principles of the Dandelion experiment.}
	\label{fig:principles}
\end{figure}

In this paper, we present Dandelion, a directional dish antenna experiment being installed at the LPSC (Grenoble, France) scheduled by the end of 2023. The DP-induced photons are detected on a 2D-matrix of Lumped Element Kinetic Inductance Detectors\cite{Day2003, Roesch2011}, hereafter referred as KIDs, cooled down to $150~\rm{mK}$ by means of a cryostat. The working principles of the Dandelion experiment are presented in Figure~\ref{fig:principles}. Dandelion benefits from an extensive experience in the detection of meV photons with KIDs from the experiences NIKA2 \cite{Perotto2019}, KISS \cite{Fasano2019}, and CONCERTO \cite{CONCERTO2020} and uses a matrix of KIDs and a cryostat initially developed for astrophysical purposes. Theoretical motivation for meV DP as a probe for the early universe is presented in Section~\ref{sec:theory}. The Dandelion experiment is described in Section~\ref{sec:Dandelion} with a focus on the background. Section~\ref{sec:signal} is dedicated to the modelling of the expected signal and its modulations. We derive the expected sensitivities in Section~\ref{sec:sensitivities}, both for exclusion limits and discovery potentials. 

\section{meV dark photons as a window towards the early universe}\label{sec:theory}

A population of dark photons that can account for the DM abundace can be  produced through different non-thermal mechanisms as: gravitational production of longitudinal massive vectors during inflation \cite{vectorDM, vectorDMEma, vectorDMSocha, vectorDMKolb}; parametric resonance and/or tachyonic instability due to an axionic scalar coupling \cite{vectorDMown, vectorDM1, vectorDM2}; production from the dark Higgs dynamics \cite{vectorDMSato,vectorDMRedi}, or from the cosmic strings associated with the spontaneous breaking of the Dark $U(1)$ symmetry \cite{vectorDMCS1, vectorDMCS2}. All of them would allow for a dark matter candidate in the meV range for the appropriate choices of model parameters, but with different cosmological implications. As an example we focus on inflationary production, comparing that of longitudinal and transverse DPs.     

Inflationary production of longitudinal modes requires the DP to be a light  massive field during inflation, i.e. $0 < m_X < H_I$, $H_I$ being the typical inflationary scale. Assuming instant or fast reheating, i.e., $m_X < H_{RH}$, $H_{RH}$ being the Hubble parameter at the end of reheating, their present abundance $\Omega_L$ normalized by the DM abundance today $\Omega_{DM}$ is given by \cite{vectorDM}:
\be
\frac{\Omega_L}{\Omega_{DM}}\simeq \sqrt{\frac{m_X}{\mev}}\left(\frac{H_I}{2.7 \times 10^{13}\,\gev} \right)^2 \,, \label{OmegaL}
\ee
where the subscript ``$L$'' refers to the longitudinal polarization. On the other hand, in the case of a late reheating, i.e., $m_X > H_{RH}$, the abundance depends on the effective equation of state during reheating $w$ \cite{vectorDMEma,vectorDMSocha,vectorDMKolb}:  
\be
\frac{\Omega_L}{\Omega_{DM}}\simeq \left(\frac{m_X}{0.33~\gev}\right)^{2w/(1+w)}\left(\frac{H_I}{3.3 \times 10^{10}\,\gev} \right)^{2} \left(\frac{2.1\times 10^{9} ~\gev}{T_{RH}} \right)^{(3 w -1)/(1+w)} \,, \label{OmegaLRH} 
\ee
where $T_{RH}$ is the reheating temperature\footnote{We have used $H_{RH} = (\pi^2 g_*/90)^{1/2}T_{RH}^2/m_P$, with $g_*=100$.}, and one recovers Eq. \eqref{OmegaL} with $w=1/3$. For  $m_X \simeq 1$ meV, reheating can affect the final abundance only for temperatures $T_{RH} \lesssim 840 ~ \gev$. In general, for a meV DP, this production mechanism requires large scale inflation, $H_I \sim O(10^{13}- 10^{14})$ GeV, in order to explain the present DM abundance, which in turn may be in tension with the current limit on the tensor-to-scalar ratio \cite{tensorlimit} $r\simeq 0.032$, and the upper bound it sets on the scale of inflation $H_I \simeq 4.44 \times 10^{13}$ GeV. For example for a matter dominated late reheating, $w=0$, the final abundance is independent of the dark photon mass and given by:
\be
\frac{\Omega_L(w=0)}{\Omega_{DM}} \simeq  \left(\frac{H_I}{5.2 \times 10^{13}\,\gev} \right)^{2} \left(\frac{T_{RH}}{840~\gev} \right) \,. 
\label{OmegaLTRH} 
\ee
Having $\Omega_L = \Omega_{DM}$ then implies $r \simeq 0.04$, already excluded. Instant reheating or $m_X < T_{RH}$, i.e. $w=1/3$, leads to $r \simeq 0.01$, while an effective $w =1$ requires $H_I \simeq 1.2 \times 10^{13}$ GeV and then $r \simeq 2\times 10^{-3}$. Therefore, this mechanism for longitudinal DP production implies a tensor-to-scalar ratio that could be determined by future experiments \cite{Simons, CMBS4, LiteBIRD}.  

On the other hand, inflationary production of transverse modes can be efficient
for smaller values of the inflationary scale, but requires an extra coupling of the inflaton field to the DP:
\be
   {\cal L}_{\rm inf} = -\frac{\alpha_V}{f} \phi X_{\mu \nu} {\tilde X}^{\mu \nu} \,,
\ee
where $\phi$ is the inflaton field, $f$ the scale of symmetry breaking, and ${\tilde F}^{\mu \nu}= \epsilon^{\mu \nu \alpha \beta} F_{\alpha \beta}/2$. This kind of couplings leads to the tachyonic production of one of the transverse components of the DP. The final abundance $\Omega_T$ depends on the value of the coupling $\alpha_V m_P/f = \sqrt{2} \xi_{end}$, and the details of the reheating period after inflation \cite{vectorDMown, vectorDMown2}:
\be
\frac{\Omega_T}{\Omega_{DM}} \simeq 2\times 10^{-13}\left(\frac{m}{\mev}\right)\left(\frac{H_{end}}{10^{12}\,\gev}\right)^{3/2} \left(\frac{10^{-1}}{\epsilon_R}\right)^3 \frac{e^{2 \pi \xi_{end}}}{\xi_{end}^3}\,, \label{OmegaT}
\ee
where now the subscript ``$T$'' refers to transverse modes, and the parameter $\epsilon_R= H_{RH}/H_{end}$ models the decrease in the energy density during reheating. The small pre-factor in Eq.\eqref{OmegaT} can be compensated by an efficient tachyonic production, i.e., a relative large value $\xi_{end} \gtrsim O(6)$. For example we get $\Omega_T = \Omega_{DM}$ with $\xi_{end} \simeq 6.5$, $H_{end} \simeq 1.9 \times 10^{10}$ GeV, and an efficient reheating $\epsilon_R =0.1$.

The mechanism for transverse DP production works even in the massless limit during inflation, the symmetry being broken later. But either case, massive or massless during inflation,  we have to impose that indeed the DP becomes non-relativistic before matter-radiation equality \cite{vectorDMown, vectorDMown2}, which translates into an upper bound for the Hubble parameter at the end of inflation:
\be
H_{end} \lesssim 1.3 \times 10^{12} ~ \epsilon_R^2 \left(\frac{m_X}{\mev}\right)^2 ~ \gev \,. \label{Hendlimit}
\ee
Therefore this production mechanism for an meV DP leads to a negligible tensor-to-scalar ratio, $r \lesssim O(10^{-5})$, contrary to the inflationary longitudinal production. 

On the other hand, transverse DP modes will source the tensor components of tensor perturbations after inflation ends. And given that tachyonic production of transverse DP only efficiently excites one of the transverse components of the DP, 
they lead to the production of parity violating gravitational waves (GW) \cite{DPGW1, DPGW2, DPGW3,DPGW}. Whether or not this GW spectrum can be detected by present and near future experiments is still an open question. Large scale inflationary models, say $H_{end} \simeq O(10^{13})$ GeV, typically give rise to a GW spectrum peaked at large frequencies, $O({\rm MHz})$, outside present capabilities of detection. Inflationary models with lower $H_{end}$ may lead to a smaller frequency peak, but also to smaller amplitudes. 

To summarize, as an example of possible production mechanism for a DP with a mass near the meV, we have compared the inflationary production of longitudinal and transverse photons. Once we fix the mass parameter, these mechanisms are mutually excluded, and we cannot have simultaneously both polarizations at production in our DM population. They will be also related to different cosmological signatures: longitudinal production leads to a tensor-to-scalar ratio within the reach of near future proposed missions, whereas this signature is negligible for transverse DP; but the latter may induce a signal in the GW spectrum at much larger frequencies. We will show in Section~\ref{sec:sensitivities} that Dandelion could determine the polarization of the DP, hence allowing to rule out one of the two inflationary production mechanisms. However, cosmological evolution of our population of DPs may lead to the conversion of longitudinal-transverse polarizations, giving rise to a random polarized background of DPs today. To study how efficient might be the conversion, depending on the production mechanism, is beyond the scope of this work, and we will just consider the two limiting cases in the following: either a \textit{fixed} polarization, or a \textit{random} polarization.

\section{The Dandelion experiment}\label{sec:Dandelion}

Dandelion stands for DArk photoN DirEctionaL detectION. Just as the pappus on a dandelion flies away according to the direction of the wind, Dandelion could trace its measurements back to the direction of the DP wind due to the motion of the Earth through the DM halo.

The conversion surface between a DP and a standard photon is a spherical aluminium mirror. While a large surface would be suitable for improving the expected DP-induced photon flux on the detector, we here operate a prototype to experimentally validate the strategy of detection. We consider a spherical mirror with a curvature radius $R=5~\rm{m}$ and an aperture diameter of $D=50~\rm{cm}$. For a spherical mirror, the incident photons parallel to the optical axis converge to $R/2$, as it is the case for astrophysical measurements, whereas the photons emitted perpendicularly to the surface focus at $R$. In other words, the DP-induced photons and the reflected ambient background photons do not converge to the same positions. We thus place the detector at a distance $R$ from the mirror. The experiment is located at Grenoble, France, at a latitude of $45.1^\circ$ and a longitude of $5.7^\circ$. The mirror is facing the direction North-West with an angle of $48^\circ$ with respect to the North. These coordinates will play a role in Section~\ref{sec:signal} to model the directional signal. 

\begin{figure}[t]
	\centering
	\begin{minipage}{0.55\linewidth}
		\includegraphics[width=\linewidth]{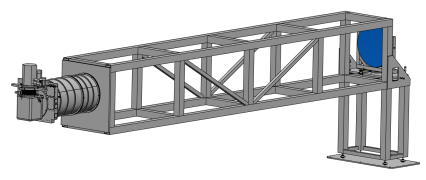}
	\end{minipage}
	\hfill
	\begin{minipage}{0.38\linewidth}
		\includegraphics[width=0.8\linewidth]{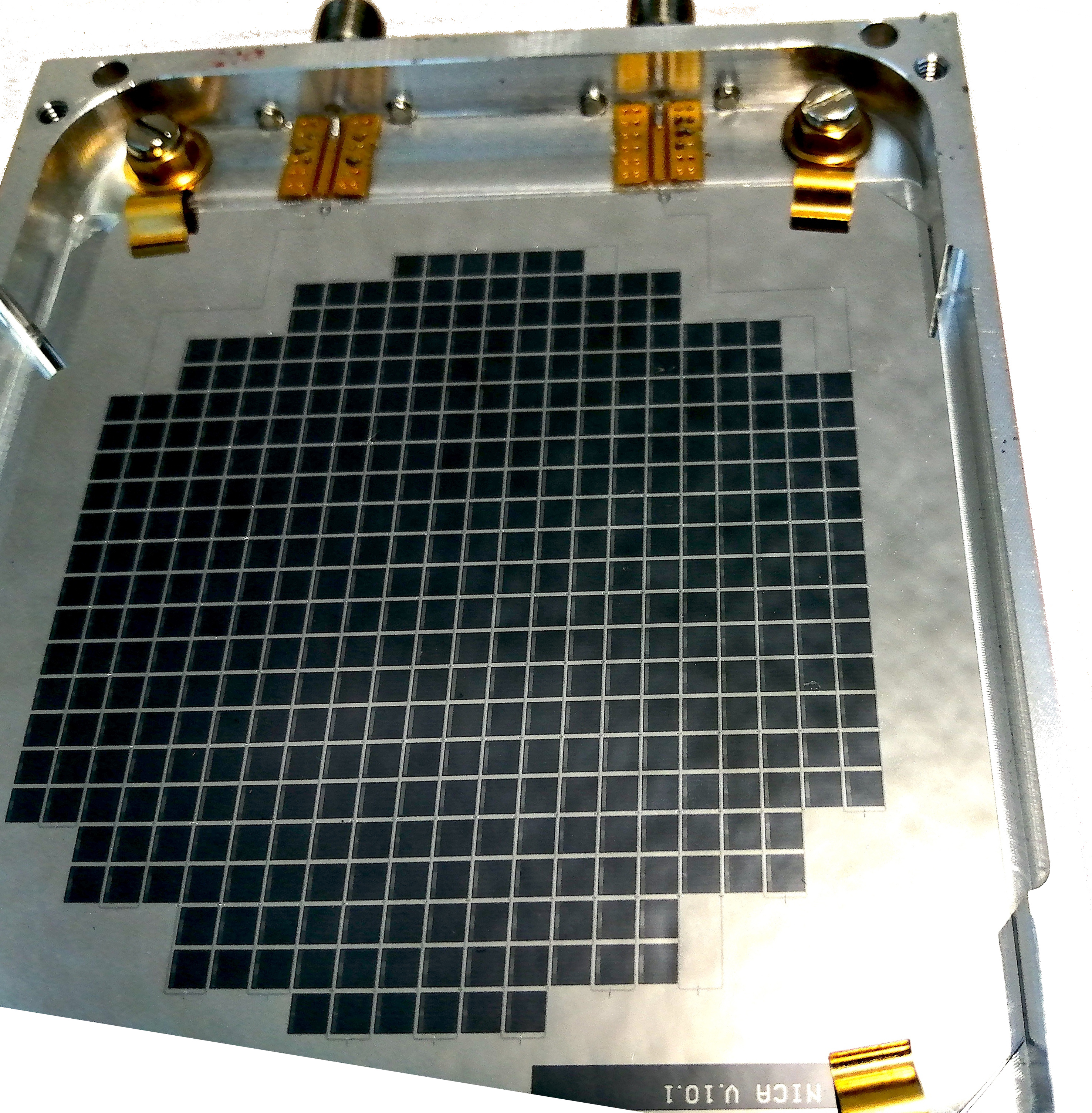}
	\end{minipage}
	\caption{Left: Computer-Aided Design (CAD) of the experimental setup. The cryostat hosting the detector is tight to the mirror by a mechanical structure. Right: the mounted matrix made of $418$ LEKID pixels.}
	\label{fig:setup}
\end{figure}

The detector consists of a matrix of KIDs that forms a disk of diameter $6.5~\rm{cm}$ made of $418$ pixels of dimensions $2.8\times2.8~\rm{mm^2}$. This matrix, the cryostat to cool it down to $150~\rm{mK}$, as well as the related acquisition soft were previously developed for other research projects \cite{AndreaHDR}. The detection efficiency of the KIDs, $\eta(E_\gamma)$, has been measured; it shows a maximum at $250~\rm{GHz}$ (corresponding to $1~\rm{meV}$) and exceeds $50\%$ between $200$ and $300~\rm{GHz}$. Some optics ensure an optimized focus on the KIDs. The alignment between the detector and the mirror is controlled every day by means of an optical laser located at the center of the mirror in which a hole has been drilled. To reduce alignment uncertainties, the mirror is mechanically tight to the cryostat. Figure~\ref{fig:setup} presents a CAD model of the setup and a picture of the matrix of KIDs. 

\begin{figure}[t]
    \centering
	\includegraphics[width=0.8\linewidth]{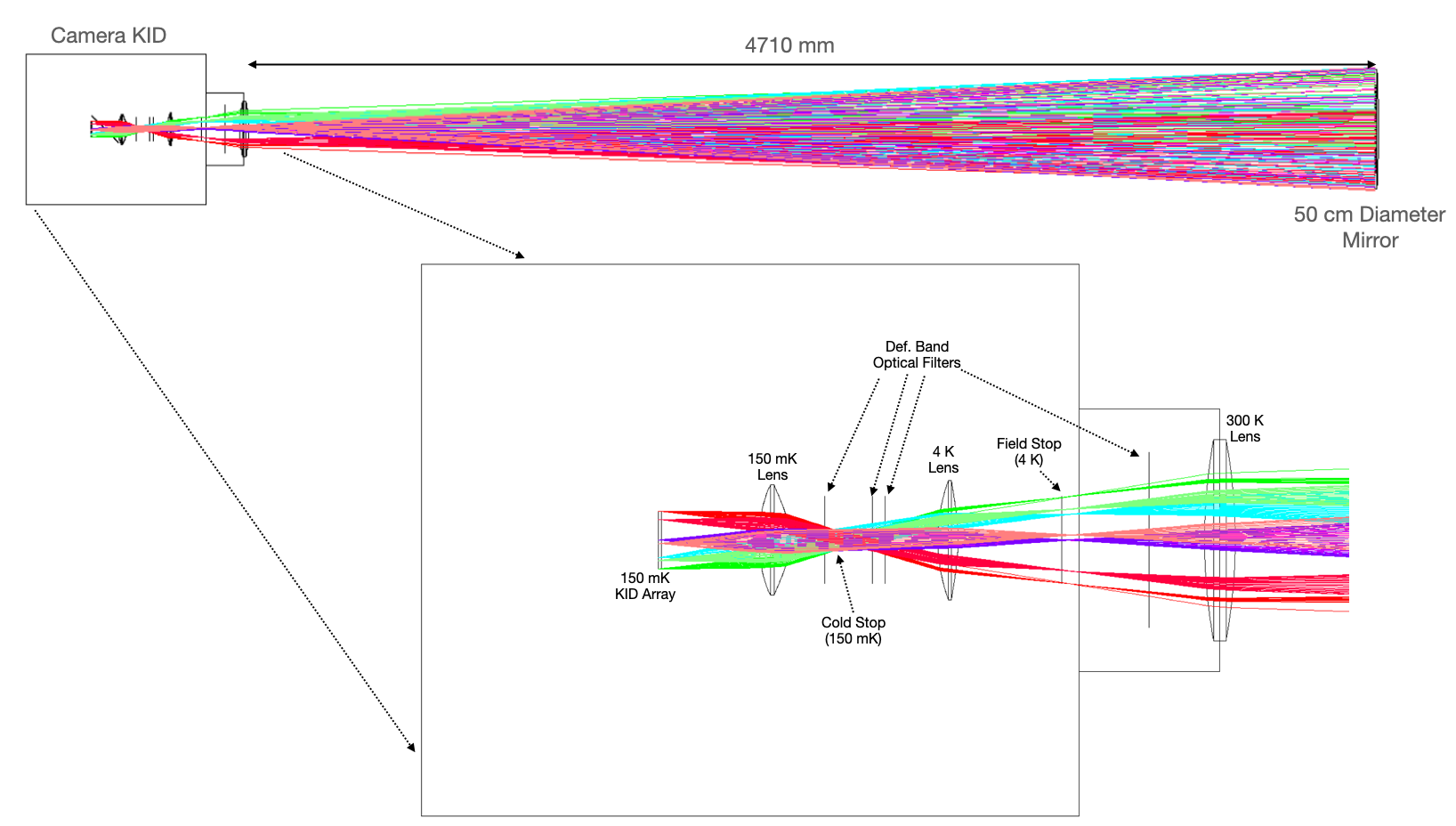}
	\caption{Dandelion optical design (Zemax software) with details of the cold optics inside the cryostat. Inverse ray tracing simulation from detectors to the $50~\rm{cm}$ diameter mirror. Detectors illumination is constrained inside the mirror in order to minimize the thermal loading.}
	\label{fig:Zemax}
\end{figure}

The main source of background is the thermal background. The background contribution is minimized by the optics placed in front of the matrix of KIDs such as the input pupil is smaller than the mirror at a distance of $R$, as detailed in Figure~\ref{fig:Zemax}. The reflected signal on the surface of the mirror is then at $150~\rm{mK}$, \textit{i.e.} the cryostat's temperature. From Planck's law, one can derive the thermal background, from emission and reflection, in one single pixel of the detector and thus determine the optical Noise Equivalent Power (NEP) defined as the signal power required to get a unit signal-to-noise ratio (SNR) in one Hertz bandwidth. We theoretically obtain a mean NEP of $3\times 10^{-17}~\rm{W/\sqrt{Hz}}$ on any of the $418$ pixels. Sources of $1/f$-like noise are suppressed by a synchronous modulation technique: the mirror is tilted by $5~\rm{mrad}$ at a frequency of $1~\rm{Hz}$ around its vertical axis (perpendicular to the main axis of the mirror) by means of a motor, resulting in a shift of $2.5~\rm{cm}$ of the expected signal spot. Such a shift is sufficient to move the signal out of the central pixels while letting it focused on some other pixels of the matrix of KIDs. The modulation of the expected signal over the pixels happens at a fast enough frequency so the background can be considered constant, hence leading to a continuous measurement of the noise for each pixel which is then subtracted for extracting a signal.  

\section{Signal modelling}\label{sec:signal}

In this section, we investigate the features of the expected signal and its spatial and intensity modulations. Some compromises must be determined in the design of the experiment to optimize the detection efficiency. For instance, a larger diameter for the mirror increases the power of the signal while it deteriorates the spatial resolution  depending on where the DP-induced photons are emitted on the mirror \cite{Jaeckel2013}. Following \cite{Jaeckel2015}, we obtain that such optical aberrations are negligible for the Dandelion geometry since they result in a maximal spread of the signal of $0.3~\rm{mm}$. 

\subsection{Diffraction}

The main limitation for a dish antenna searching for DPs with masses typically below $100~\rm{meV}$ comes from the diffraction. After the emission of a DP-induced photon, the mirror acts as a source of secondary wavelets which interfere with the primary  wave. According to the Huygens-Fresnel principle, the electric field on the matrix of KIDs, $E(x,y,R)$, is related to the electric field emitted at the surface of the mirror, $E(x', y', 0)$, by \cite{Goodman2005}:

\begin{equation}
	E(x,y,R) = \frac{R}{i\lambda}\,\int_{\rm{mirr.}} dx'\,dy'\,E(x', y', 0)\,\frac{e^{-ikr}}{r^2}\,,
		\label{eq:HyugensFresnel}
\end{equation}
where $k=m/(\hbar c)$ is the wavenumber and $r$ is the distance between a point with coordinate $(x', y')$ on the mirror and a point with coordinate $(x,y)$ on the readout plane, which can be expressed as:
\begin{equation}
	r = R \bigg\lbrace 1 + \frac{(x-x')^2 + (y-y')^2}{2R^2} - \frac{x'^2 + y'^2}{2R^2} + o\Big( \big(\frac{x}{R}\big)^4 \Big) \bigg\rbrace\,.
\end{equation}

Since we are in a near-field configuration, we apply the Fresnel approximation to obtain:

\begin{equation}
	E(x,y,R) \simeq \frac{e^{-ikR}}{i\lambda R}\int_{\rm{mirr.}} dx'dy'E(x', y', 0)\,\exp\bigg( -\frac{ik}{2R}\Big((x-x')^2 + (y-y')^2 \Big) \Bigg) \exp\bigg(\frac{ik(x'^2 + y'^2)}{2R} \Bigg) \,.
		\label{eq:Fresnel}
\end{equation}

\begin{figure}[t]
	\centering
	\begin{minipage}{0.49\linewidth}
		\includegraphics[width=\linewidth]{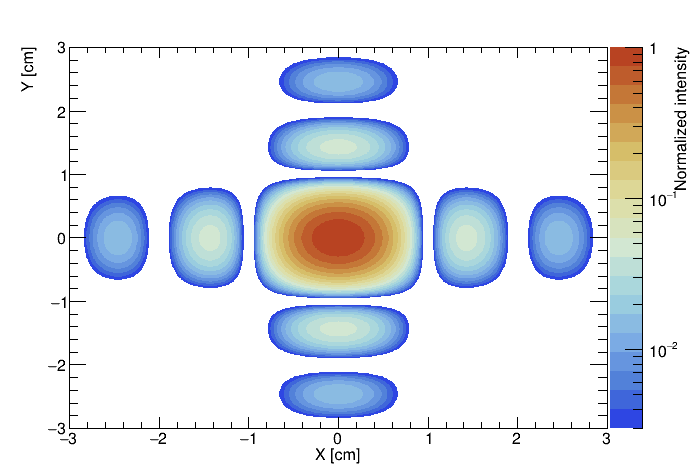}
	\end{minipage}
	\hfill
	\begin{minipage}{0.49\linewidth}
		\includegraphics[width=\linewidth, height=5.2cm]{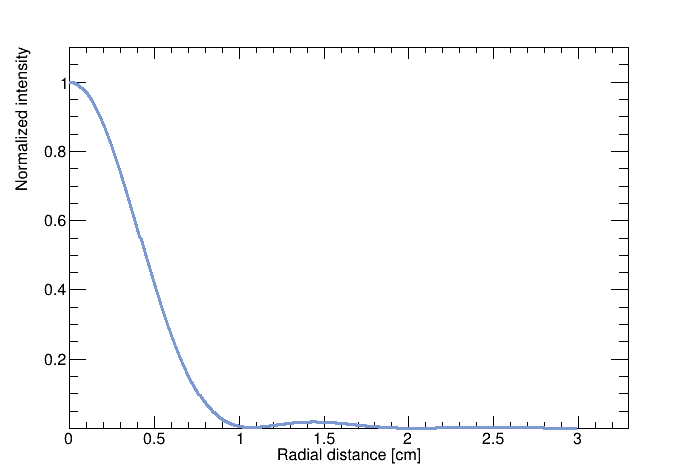}
	\end{minipage}
	\caption{The diffraction of the signal at the readout plane. Left: representation 2D. Right: projection along the radial component.}
	\label{fig:diffraction}
\end{figure}

The last exponential term is an additional phase compared to the expression of the Fresnel diffraction in textbooks that treat the case of a circular aperture whereas here we consider a spherical surface. This phase significantly reduces the influence of the diffraction. The spread of the signal due to the diffraction is represented in Figure~\ref{fig:diffraction} from a numerical computation of Eq.~\eqref{eq:Fresnel} and is comparable to the results of the Zemax simulation. This spread limits the spatial resolution of the detection: $1/3$ of the signal is contained in the 5 central pixels; and $2/3$ of the signal in the 13 central pixels. 

\subsection{Spatial modulation: the directional signature}

Let us now derive the spatial modulation of the signal which will be compared in Section~\ref{sec:sensitivities} to the spatial resolution to evaluate the discovery potential of the detector. The spatial modulation is related to the out-of-normal photon emission with an angle $\psi = |\vec{p}_{||}(t)|/m$ that we determine in two steps. First, we express $-\vec{v}_{\rm{lab}}$, the DP velocity vector in the mirror frame, accounting for multiple motions: the galactic rotation at the position of the Sun, the peculiar velocity of the Sun in the galaxy, the revolution of the Earth around the Sun, and the rotation of the Earth. Second, we project $-\vec{v}_{\rm{lab}}$ along the tangent to the mirror. We deeply rely on \cite{Bozorgnia2011} for this analysis. Appendix~\ref{app:motions} extrapolates some equations of their paper to our configuration and we provide the expression for the spatial modulation in Eq.~\eqref{eq:offsetKIDs}. 

\begin{figure}[t]
	\centering
	\includegraphics[width=0.7\linewidth]{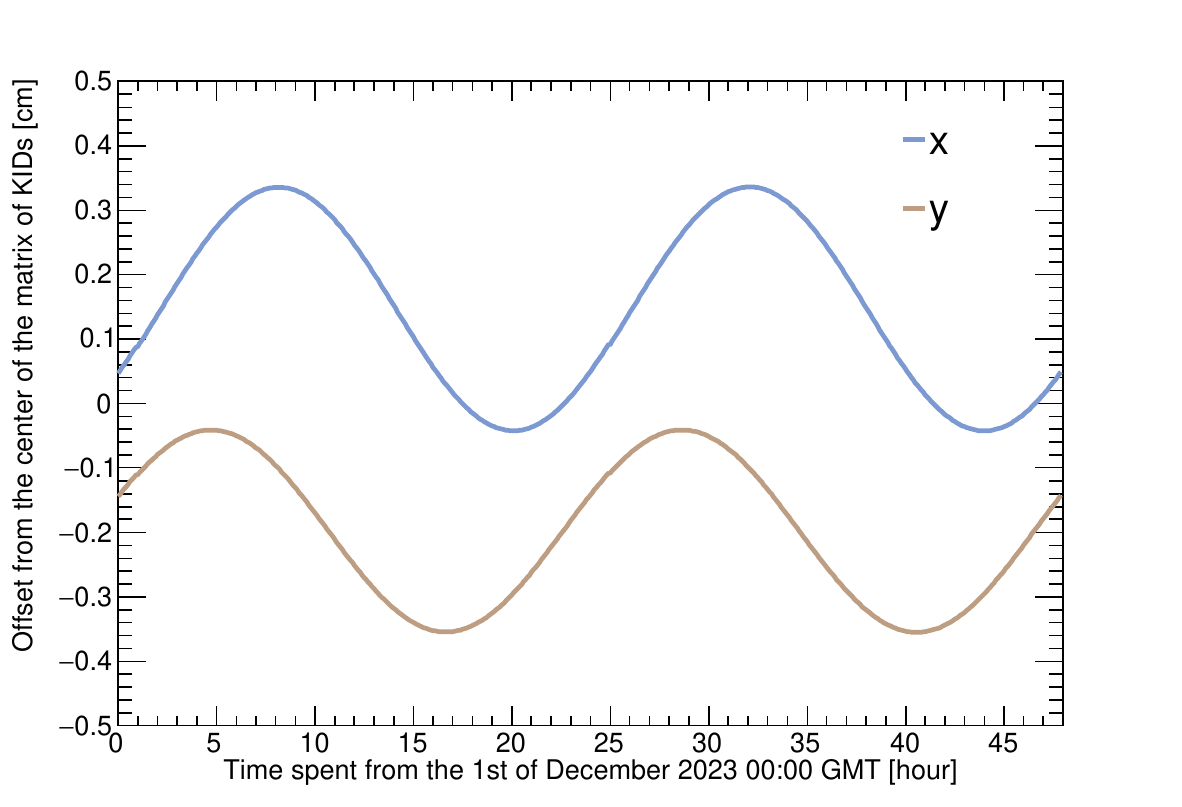}
	\caption{The expected spatial modulation for a 48h-long measurement starting on the 1st of December 2023. Based on the modelling of \cite{Bozorgnia2011}.}
	\label{fig:offset}
\end{figure}

A typical spatial modulation is presented in Figure~\ref{fig:offset} which exhibits the tendencies that are observed for any input day: the offset variations in X and Y cover between one et two pixels; the signal modulates with a period of one day, reaching an extrema every 12h; there is a delay of about 2h between the extrema in X and in Y. These features are intrinsically derived from the hypothetical existence of a DM halo surrounding the galaxy, and they cannot be reproduced by the background. While the detection sensitivity is studied in Section~\ref{sec:sensitivities}, we can already say that the power induced on a single pixel varies up to a factor of 4 between two extrema once accounting for the diffraction.

\subsection{Intensity modulation}\label{subsec:intMod}

The detectable signal modulates in intensity proportionally to the cosine squared of the angle between the polarization of the DP and the tangent of the mirror, \textit{i.e.} the $\alpha(t)$ angle of Eq.~\eqref{eq:power}. For the data analysis, we will integrate the measured signal over a period $T$, so the quantity of interest becomes $\langle\cos^2\alpha(t)\rangle_T$. 

An important contribution to the modelling of $\alpha(t)$ is presented in \cite{Caputo2021}. We decide however to approach the modelling from another point of view based on the description of the relative motion of the mirror in the galactic halo, as detailed at the end of Appendix~\ref{app:motions}. This approach simplifies the connection between the spatial and intensity modulations in our work. Moreover, we express the DP polarization in the galactic frame whereas it is expressed in the geocentric frame in \cite{Caputo2021} as an approximation. 

\begin{figure}[t]
	\centering
	\begin{minipage}{0.49\linewidth}
		\includegraphics[width=\linewidth]{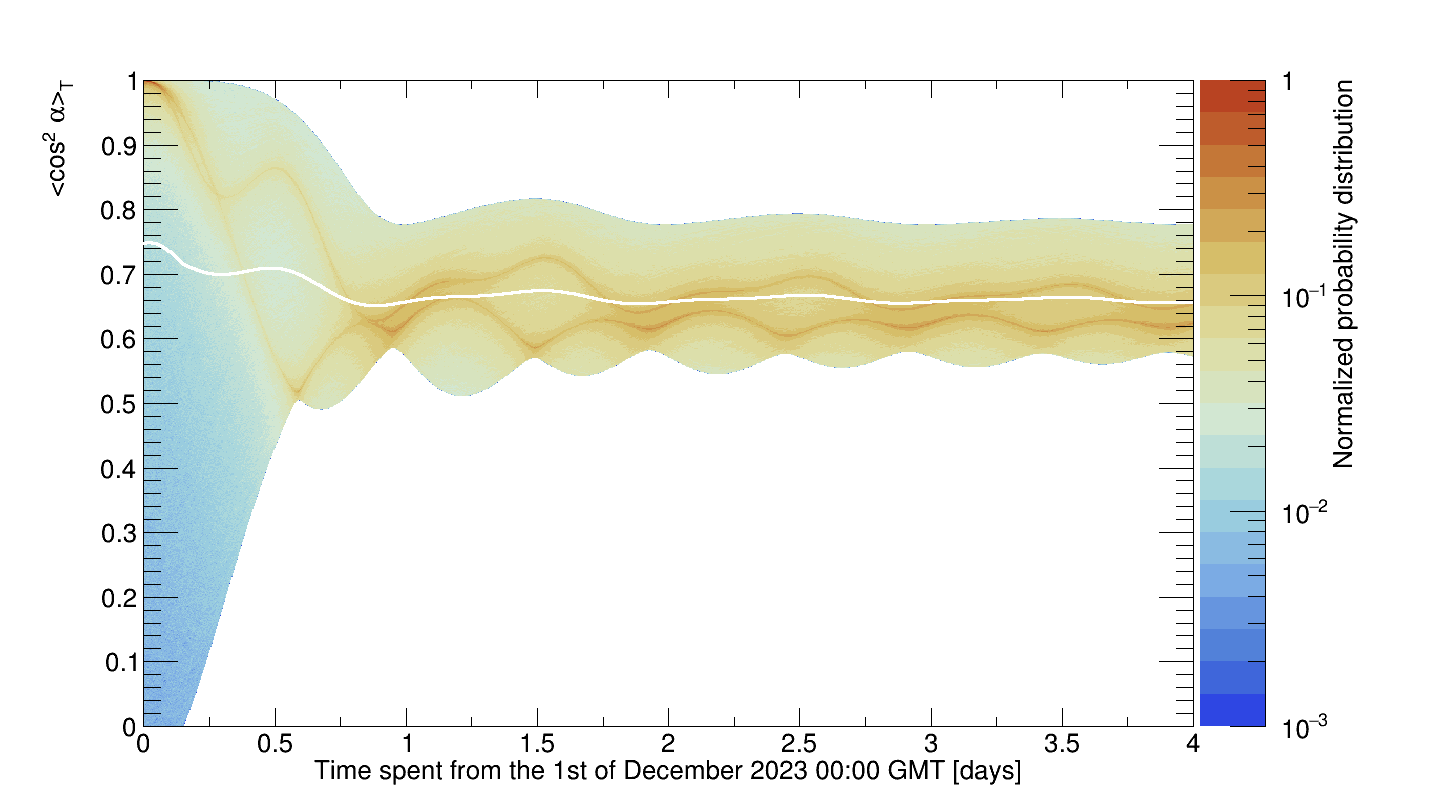}
	\end{minipage}
	\hfill
	\begin{minipage}{0.49\linewidth}
		\includegraphics[width=\linewidth]{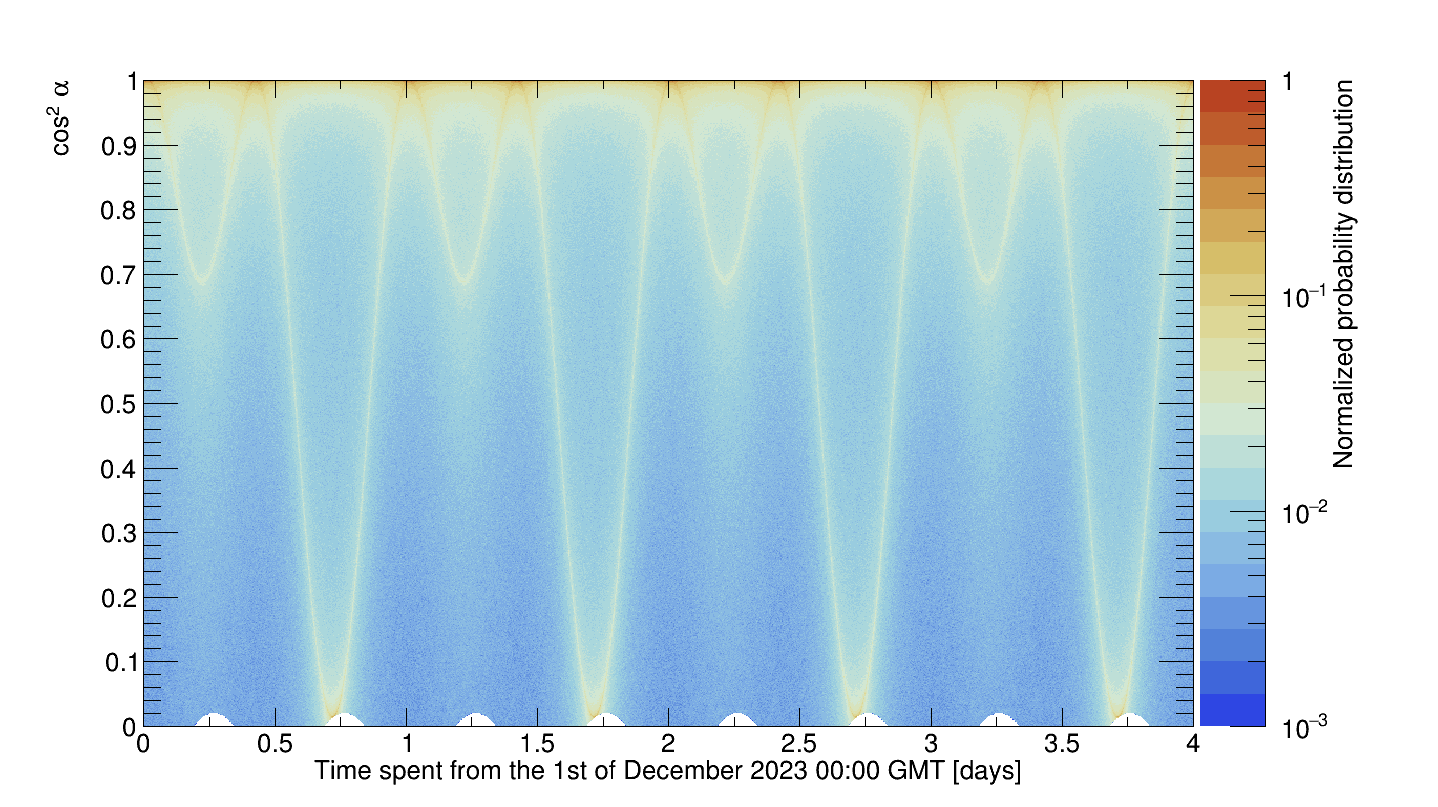}
	\end{minipage}
	\caption{Time evolution of the probability distribution of the time-averaged $\langle\cos^2\alpha(t)\rangle_T$ (left) and of $\cos^2\alpha(t)$ (right), where $\alpha(t)$ is the angle between the DP polarization and the tangent of the mirror. The probability distributions are scaled such that their maxima are equal to unity. The plots are constructed from $10^5$ DP polarizations that are randomly drawn from an isotropic distribution in the galactic frame. In the left plot, the distribution of the mean value at each time is represented as a white curve.}
	\label{fig:polarization}
\end{figure}

Since the DP polarization is unknown, we will consider the two scenarios presented in Section~\ref{sec:theory}: the \textit{random} polarization case and the \textit{fixed} polarization case. Figure~\ref{fig:polarization} presents the time evolution of $\langle\cos^2\alpha(t)\rangle_T$ and $\cos^2\alpha(t)$ for our geometry based on $10^5$ random DP polarizations and computed from Eq.~\eqref{eq:cos2}. In the figure, the modelling starts on the 1st of December 2023. By systematically varying the initial time, we can establish the following points:
\begin{itemize}
	\item While the most probable values of the distribution of $\langle\cos^2\alpha(t)\rangle_T$ depend on the initial time, the mean value only feebly fluctuates. One can conservatively assume $\langle\cos^2\alpha(t)\rangle_T \simeq 2/3$ for all $T$ larger than one hour in the \textit{random} scenario.
	\item The instantaneous value of $\cos^2\alpha(t)$ presents two minima and two maxima in $24~\rm{h}$. The maxima always correspond to $\cos^2\alpha(t) = 1$.
	\item In the \textit{random} scenario, the values of the minima feebly evolve depending on the initial time: one near $0$ and the other near $0.7$.
	\item In the \textit{fixed} scenario, the values of the minima significantly depend on the polarization components.
\end{itemize}

The intensity modulation presents a detectable signature that, in case of detection, could enable the identification of the polarization of the DP. However, since the polarization of the DP is unknown, the intensity modulation is not a robust enough signature for the directional detection compared to the spatial modulation.

\section{Expected sensitivities}\label{sec:sensitivities}

We now have all the required elements to evaluate the power in Eq.~\eqref{eq:power}. We use $\rho_{\rm{CDM}} = 0.45~\rm{GeV\,cm^{-3}}$ for consistency with other DP searches \cite{Caputo2021}. The signal spread $I(x,y,t)$ embeds both the diffraction and the spatial modulation and is normalized such that $\int dx dy\,I(x,y,t) = 1$ for each $t$. This quantity and $\cos^2\alpha(t)$ are computed numerically to account for the detector's motion through the galactic halo.

We rely on a simple and conservative approach for the derivation of the sensitivities. Finer limits could be obtained in the future, for instance based on dedicated methods for directional detection  \cite{Billard2010, Billard2010bis, Mayet2016}. For an integration time $T$, the DP-induced signal on the pixel of coordinate $(i,j)$ is given by:
\begin{equation}
	S_{ij}(E_\gamma) = \frac{1}{T}\int_0^Tdt \int_{\rm{pixel}}dx\,dy\,P(x, y, t;E_\gamma)\,,
\end{equation}
where the spatial integral only runs over the dimension of the pixel $(i,j)$. The mirror is tilted by $5~\rm{mrad}$ at a frequency of $1~\rm{Hz}$ 
so the tilt is small enough to keep the expected DP signal inside the matrix of the $418$ KIDs. In other words, the entire experimental time is used both for signal and background measurements. Since this paper precedes the data acquisition, we assume that all pixels experience a constant noise equal to the NEP, so the background on the pixel $(i,j)$ after an integration time $T$ is:

\begin{equation}
	B_{ij} \simeq \frac{\rm{NEP}}{\sqrt{T}}\,.
\end{equation}

We define a region of interest (ROI) of the 13 pixels receiving the maximal expected power. Increasing or reducing the diameter of the ROI, up to a factor of 2, has negligible influence on the results. A DP-like signal would be detected if the following requirement is fulfilled:
\begin{equation}
\frac{\sum_{i,j}S_{ij}(E_\gamma)}{\sum_{i,j}B_{ij}} > 5 \,,
\end{equation}
where $(i,j)$ are located inside the ROI. This condition defines the Dandelion exclusion sensitivity.

The discovery potential requires an analysis of the spatial modulation. We define two signal regions corresponding to the extreme deciles of the radius of the spatial modulation: $S_{ij}^{10}$ integrates the power in pixel $(i,j)$ only for the times that fulfil the condition $r(t) < 0.1(r_{\rm{max}} - r_{\rm{min}}) + r_{\rm{min}}$, and $S_{ij}^{90}$ only for $r(t) > 0.9(r_{\rm{max}} - r_{\rm{min}}) + r_{\rm{min}}$. Similarly, $B_{ij}^{10}$ and $B_{ij}^{90}$ are the background measurements during the same time periods. For determining the discovery potential, we only keep the pixels in which the SNR>5 and we require an unambiguous signal difference between the two extreme deciles:
\begin{equation}
	\sum_{i^*, j^*}\bigg|1 - \frac{S_{i^*j^*}^{90}}{S_{i^*j^*}^{10}}\bigg| > 5 \hspace*{0.6cm} \rm{with}~(i^*,j^*)~\rm{such~that} \hspace*{0.3cm} \frac{S^{10}_{i^*j^*}}{B^{10}_{i^*j^*}} > 5 \hspace*{0.3cm}\rm{and}\hspace*{0.3cm} \frac{S^{90}_{i^*j^*}}{B^{90}_{i^*j^*}} > 5\,.
\end{equation}

If DPs are detected, we could go one step further by  determining if the polarization is random or fixed. The $S^{90}_{i^*j^*}$ signal covers a time period of about 6 hours in which the $\cos^2\alpha(t)$ goes from one maximum to one minimum.  We cut this period in 5 timesteps, noted $ts$, such that in the \textit{random} scenario $\langle\cos^2\alpha(t)\rangle_{ts} \simeq 2/3$. In the \textit{fixed} scenario, the evolution of $\langle\cos^2\alpha(t)\rangle_{ts}$ depends on the polarization although in most cases it varies by about a factor of 5 so the intensity modulation might be observed from one timestep to another. A conservative limit is placed by requiring that the signal difference between two timesteps exceeds twice the background. This condition is added to the requirements for the discovery of the DP.

\begin{figure}[t]
	\centering
	\includegraphics[width=0.7\linewidth]{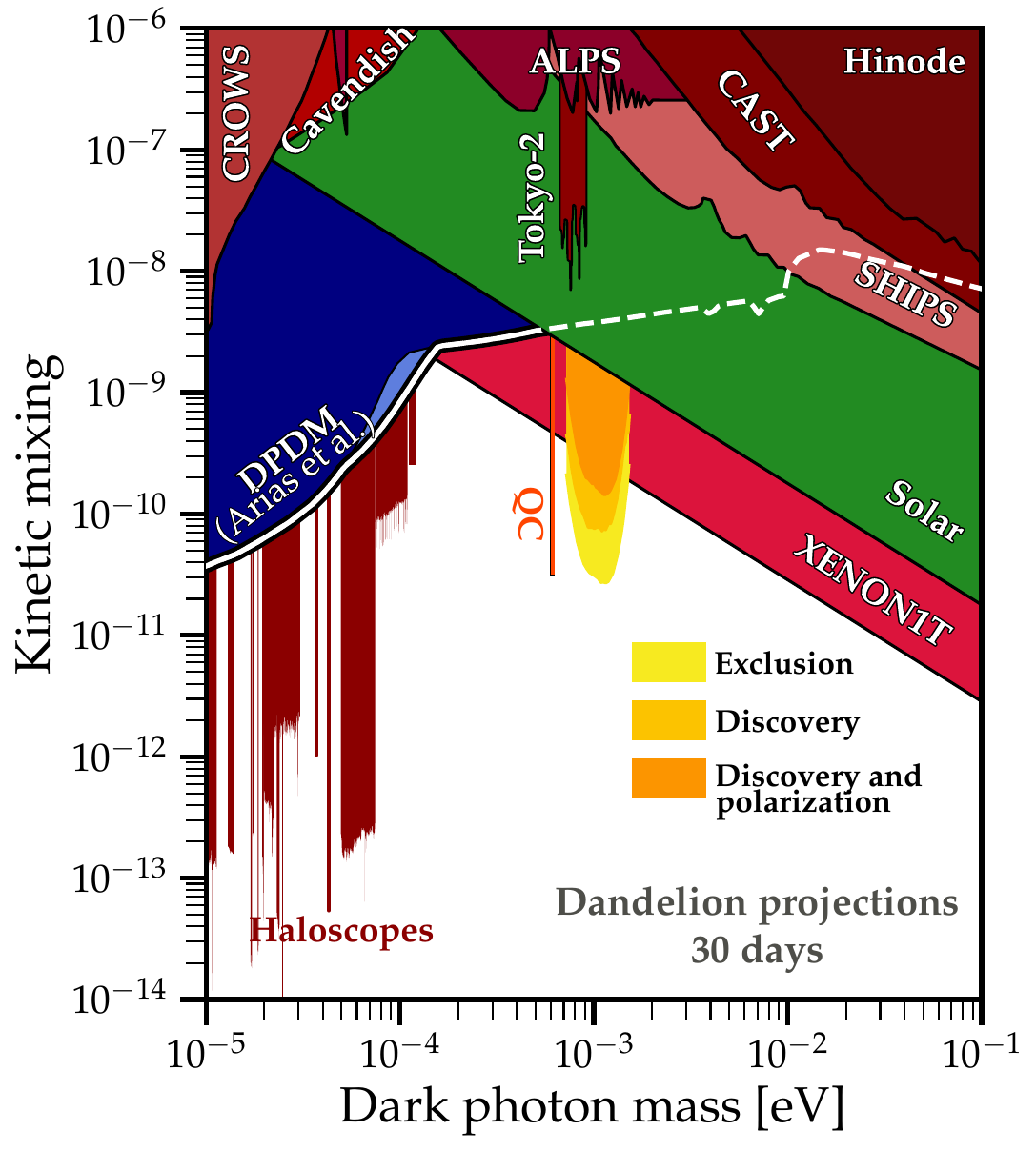}
	\caption{The expected sensitivities for the Dandelion prototype running for 30 days. We show the exclusion limits, the discovery potential, and the discovery potential discriminating between the polarization scenarios. Figure adapted from the Github of Ciaran O'Hare \cite{Cajohare}.}
	\label{fig:limits}
\end{figure}

The projected sensitivities of Dandelion for a test campaign of 30 days are presented in Figure~\ref{fig:limits}. The prototype described in this paper would improve by more than one order of magnitude the existing limits around the meV DP mass. It would also provide an unprecedented discovery potential and could discriminate between two polarization scenarios. Dandelion would probe a significantly larger parameter space than Tokyo-2 \cite{Knirck2018} which also exploits the dish antenna principle without the directional information.   

\section{Conclusion}\label{sec:conclu}

The directional detection of a DM particle would represent a momentous and exciting discovery. The dark photon, besides being a suitable DM candidate, can serve as a portal to a dark sector. While the mass of a DP can cover more than twenty decades, the meV region is of particular interest since it has few experimental constraints and that the polarization of a meV DP provides information about the primordial universe. 

In this paper, we present the Dandelion experiment that is currently being installed at the LPSC (Grenoble, France). Dandelion is an experiment that benefits from an already operational matrix of KIDs, cooled down to $150~\rm{mK}$ by a cryostat, to detect DP-induced photons. We have shown that the experiment would be sensitive both to the spatial and intensity modulations of the signal related to the motion of the detector through the galactic halo. A small Dandelion prototype and a testing measurement campaign of 30 days would improve by more than one order of magnitude the exclusion limits near the meV. If a signal is detected, the directional signature would check without ambiguity the galactic origin of the signal, hence demonstrating or not the discovery of a DM particle. 

The first campaign of the Dandelion prototype is planned for the end of 2023. Some near-term updates could improve by several orders of magnitude the sensitivity of the detector: increasing the measurement period, building a larger mirror, using multiple matrices of KIDs to cover a larger parameter space, and improving the statistical treatment of the data. If a signal is detected, we might determine the DP polarization from the intensity modulation but also measure it by means of a Martin-Puplett interferometer with only few modifications to the setup since it was previously used with the matrix of KIDs at LPSC. Another future upgrade relies on the use of tunable filters to reduce the probed bandpass. While the KIDs cover a bandwidth $\Delta \nu$, one can reduce the thermal background while not altering the signal, whose energy is given by the DP mass, by performing $n$ runs of $\Delta \nu /n$ bandwidths. Spring is just around the corner for the dandelion to bloom.

\acknowledgments

The authors thank Christopher Smith, Jérémie Quevillon, François-Xavier Désert and Fabrice Naraghi for the fruitful discussions at the first steps of the project. The authors are deeply grateful to Julien Bounmy and Damien Tourres for their support with the electronics. The authors acknowledge funding from the French Programme d’investissements d’avenir through the Enigmass Labex.

\appendix
\section{Mirror's motion through the galactic DP halo}\label{app:motions}

We entirely rely on \cite{Bozorgnia2011} for modelling the motion of the mirror with respect to the galactic DP halo. This appendix transposes their main results to our experimental setup. Let us first introduce some coordinate systems. We define a reference frame for the mirror with unit vectors  $(\hat{\vec{X}}, \hat{\vec{Y}}, \hat{\vec{Z}})$ where $\hat{\vec{X}}$ and $\hat{\vec{Y}}$ lie in the plane of the mirror and $\hat{\vec{Z}}$ is perpendicular to the tangent of the mirror. The origin of the coordinate system is located at the center of the mirror and we assume the coordinate system to be right-handed. 

We now define a reference frame fixed to the laboratory with unit vectors $(\hat{\mathcal{N}}, \hat{\mathcal{W}}, \hat{\mathcal{Z}})$ for which $\hat{\mathcal{N}}$ points to the North, $\hat{\mathcal{W}}$ to the West, and $\hat{\mathcal{Z}}$ to the Zenith. The relation between the mirror frame and the laboratory frame is described by the following coefficients:
\begin{eqnarray}
\hat{\mathbf{X}}&=&\alpha_X~\hat{\mathbf{\mathcal{N}}}+\beta_X~\hat{\mathbf{\mathcal{W}}}+\gamma_X~\hat{\mathbf{\mathcal{Z}}}\nonumber\\
\hat{\mathbf{Y}}&=&\alpha_Y~\hat{\mathbf{\mathcal{N}}}+\beta_Y~\hat{\mathbf{\mathcal{W}}}+\gamma_Y~\hat{\mathbf{\mathcal{Z}}}\\
\hat{\mathbf{Z}}&=&\alpha_Z~\hat{\mathbf{\mathcal{N}}}+\beta_Z~\hat{\mathbf{\mathcal{W}}}+\gamma_Z~\hat{\mathbf{\mathcal{Z}}}\nonumber
\end{eqnarray}

The main axis of the experimental setup, the $\hat{\vec{Z}}$ axis, is perpendicular to the wall of the experimental room and we approximate the mirror as a disk instead of a spherical surface. In such a case, $\hat{\vec{Y}}$ is oriented along $\hat{\mathbf{\mathcal{Z}}}$. The mirror is oriented N-W with an angle of $\theta_{NW} = 48^\circ$ from the North. We can thus approximate the coefficients:
\begin{align}
	\alpha_X &\simeq  -\sin(\theta_{NW}) & \beta_X &\simeq \cos(\theta_{NW}) & \gamma_X &\simeq 0 \nonumber\\ 
	\alpha_Y &\simeq  0 & \beta_Y &\simeq 0 & \gamma_Y &\simeq 1 	\label{eq:coefficientMirrorFrame}
\\
	\alpha_Z &\simeq \cos(\theta_{NW}) & \beta_Z &\simeq \sin(\theta_{NW}) & \gamma_Z &\simeq 0 \nonumber
\end{align}

We want to determine $\vec{v}_{||} (t)$, the DP-velocity which is parallel to the surface of the mirror as a function of time. To do so, we define $\hat{\vec{q}}$ a unit vector in the mirror frame that is used to project $-\vec{v}_{\rm{lab}}$ along any direction, $-\vec{v}_{\rm{lab}}$ being the DP velocity vector in the mirror frame. According to \cite{Bozorgnia2011}, one has:

\begin{align}
- \hat{\mathbf{q}} \cdot{\bf v}_{{\rm lab}}=&\bigg\{\bigg[\cos(t^\circ_{\rm lab})~A(t)-\sin(t^\circ_{\rm lab})~B(t)\bigg]\sin\lambda_{\rm lab}
-C(t)~\cos\lambda_{\rm lab}\bigg\}\left(\alpha_X q_X+\alpha_Y q_Y+\alpha_Z q_Z\right)\nonumber\\
&-\bigg\{\sin(t^\circ_{\rm lab})~A(t)+\cos(t^\circ_{\rm lab})~B(t)-0.465 \cos \lambda_{\rm lab}\bigg\}\left(\beta_X q_X+\beta_Y q_Y+\beta_Z q_Z\right)\nonumber\\
&-\bigg\{\bigg[\cos(t^\circ_{\rm lab})~A(t)-\sin(t^\circ_{\rm lab})~B(t)\bigg]\cos\lambda_{\rm lab}+C(t)~\sin\lambda_{\rm lab}\bigg\}\left(\gamma_X q_X+\gamma_Y q_Y+\gamma_Z q_Z\right)
\label{eq:qVlab}
\end{align}

where

\begin{align}
A(t)&=0.4927~ V_{\rm {Gal Rot}} - 1.066~\textrm{km/s} + (V_{\oplus}(\lambda(t)) {\cal A}(t)\nonumber\\
B(t)&=0.4503~ V_{\rm {Gal Rot}} + 16.56~\textrm{km/s} - (V_{\oplus}(\lambda(t)) {\cal B}(t)\nonumber\\
C(t)&=0.7445~ V_{\rm {Gal Rot}} + 7.077~\textrm{km/s} + (V_{\oplus}(\lambda(t)) {\cal C}(t)
\end{align}

\noindent and where we defined multiple quantities:
\begin{itemize}
	\item $V_{\rm {Gal Rot}}$ is the velocity of the rotation of the galaxy taken as $V_{\rm {Gal Rot}} = 220~\rm{km/s}$;
	\item $\lambda_{\rm{lab}} = 45.1^\circ$ is the latitude of the experiment;
	\item $l_{\rm{lab}} = 5.7^\circ$ is the longitude of the experiment;
	\item $t^\circ_{\rm lab}$ is the local apparent sidereal time converted into degrees. In \cite{Bozorgnia2011}, the authors parametrize this quantity as:
	\begin{equation}
		t^\circ_{\rm lab} ~=~ 101.0308 ~+~ 36000.770\,T_0 ~+~ 15.04107\,\mathrm{UT}  ~+~ l_{\rm{lab}}
	\end{equation}
	We derive alternate expressions for $T_0$ and $\mathrm{UT}$ based on the Unix time given in seconds:
	\begin{equation}
		\begin{cases}
		T_0 = 3.1688\times 10^{-10} t_{\mathrm{Unix}} ~-~ 0.4 \\
		 \mathrm{UT} = \mathrm{mod}\big(t_{\mathrm{Unix}}, 86400\big) / 3600
		\end{cases}
	\end{equation}
	\item $V_{\oplus}(\lambda(t)) = V_{\oplus}\Big[1 - e\sin\big(\lambda(t) - \lambda_0\big)\Big]$ with $V_{\oplus} = 29.8~\rm{km/s}$ the orbital speed of the Earth, $e=0.016722$ the ellipticity of the Earth's orbit, $\lambda_0 = 13^\circ + 1^\circ$ the ecliptic longitude of the orbit's minor axis, and 
	\begin{equation}
		\lambda(t) ~=~ L + \big(1^\circ.915 - 0^\circ.0048\,T_0\big)\sin g + 0^\circ.020\sin 2g
	\end{equation}
where  
\begin{equation}
	\begin{cases}
		L=281^\circ .0298 + 36000^\circ .77 T_0 + 0^\circ .04107\,\rm{UT}\\
		g=357^\circ .9258 + 35999^\circ .05 T_0 + 0^\circ .04107 \rm{UT}
	\end{cases}
\end{equation}
are, respectively, the mean longitude of the Sun corrected for aberration and the mean anomaly.
\item Finally :
\begin{align}
{\cal A}(t)=&-0.06699 \cos\beta_x \sin (\lambda(t)-\lambda_x) + 0.4927\cos\beta_y \sin (\lambda(t)-\lambda_y) -0.8676\cos\beta_z \sin (\lambda(t)-\lambda_z)\nonumber\\
{\cal B}(t)=& -0.8728\cos\beta_x \sin (\lambda(t)-\lambda_x) -0.4503 \cos\beta_y \sin (\lambda(t)-\lambda_y) -0.1883 \cos\beta_z \sin (\lambda(t)-\lambda_z)\nonumber\\
{\cal C}(t)=& -0.4835 \cos\beta_x \sin (\lambda(t)-\lambda_x) + 0.7446 \cos\beta_y \sin (\lambda(t)-\lambda_y) + 0.4602 \cos\beta_z \sin (\lambda(t)-\lambda_z)
\end{align}
where 
\begin{equation}
	\begin{cases}
	\beta_i=\Big(-5^\circ.5303,~ 59^\circ.575, ~ 29^\circ.812\Big)\\
	\lambda_i=\Big(266^\circ.141, ~-13^\circ.3485, ~ 179^\circ.3212\Big)
	\end{cases}
\end{equation}
\end{itemize}

The coordinates $\big(x(t), y(t)\big)$ at which the signal is formed on the array of KIDs, with $(0,0)$ the center of the array, can be determined by projecting $-\vec{v}_{\rm{lab}}$ to the X and Y axes of the mirror frame and multiplying by the curvature radius $R$ \cite{Jaeckel2013}. Using the coefficients of Eq.~\eqref{eq:coefficientMirrorFrame}, we get:
\begin{eqnarray}
	x(t) & \simeq & R\sin(\theta_{NW})\,\bigg\{\bigg[-\cos(t^\circ_{\rm lab})~A(t)+\sin(t^\circ_{\rm lab})~B(t)\bigg]\sin\lambda_{\rm lab}
+C(t)~\cos\lambda_{\rm lab}\bigg\} \nonumber\\
&&-R\cos (\theta_{NW})\,\bigg\{\sin(t^\circ_{\rm lab})~A(t)+\cos(t^\circ_{\rm lab})~B(t)-0.465 \cos \lambda_{\rm lab}\bigg\} \nonumber\\
	y(t) & \simeq & R\bigg\{\bigg[-\cos(t^\circ_{\rm lab})~A(t)+\sin(t^\circ_{\rm lab})~B(t)\bigg]\cos\lambda_{\rm lab}-C(t)~\sin\lambda_{\rm lab}\bigg\} \label{eq:offsetKIDs}
\end{eqnarray}

In subsection~\ref{subsec:intMod}, the angle between the polarization of the DP and the tangent of the mirror must be evaluated. Since the DP polarization is expressed in the galactic frame, one needs to transform it to the mirror frame. Once again, we rely on \cite{Bozorgnia2011} to do so. We here only give the expression for the projection of the galactic frame $(\hat{\vec{x}_g}, \hat{\vec{y}_g}, \hat{\vec{z}_g})$ along $\hat{\vec{Z}}$ that is required for our analysis:
\begin{eqnarray}
\hat{\vec{x}}_g \cdot \hat{\vec{Z}} & \simeq & -0.06699\Big\{-\cos\theta_{NW}\cos t^\circ_{\rm lab}\sin  \lambda_{\rm lab} +  \sin\theta_{NW}\sin t^\circ_{\rm lab} \Big\} \nonumber\\
&& + 0.8728\Big\{\sin\theta_{NW}\sin t^\circ_{\rm lab}\sin  \lambda_{\rm lab} +  \sin\theta_{NW}\cos t^\circ_{\rm lab} \Big\} ~~ - 0.4835\cos\theta_{NW}\cos  \lambda_{\rm lab} \nonumber\\
\hat{\vec{y}}_g \cdot \hat{\vec{Z}} & \simeq & 0.4927\Big\{-\cos\theta_{NW}\cos t^\circ_{\rm lab}\sin  \lambda_{\rm lab} +  \sin\theta_{NW}\sin t^\circ_{\rm lab} \Big\} \nonumber\\
&& + 0.4503\Big\{\sin\theta_{NW}\sin t^\circ_{\rm lab}\sin  \lambda_{\rm lab} +  \sin\theta_{NW}\cos t^\circ_{\rm lab} \Big\} ~~ + 0.7446\cos\theta_{NW}\cos  \lambda_{\rm lab} \nonumber\\
\hat{\vec{z}}_g \cdot \hat{\vec{Z}} & \simeq & -0.8676\Big\{-\cos\theta_{NW}\cos t^\circ_{\rm lab}\sin  \lambda_{\rm lab} +  \sin\theta_{NW}\sin t^\circ_{\rm lab} \Big\} \nonumber\\
&& + 0.1883\Big\{\sin\theta_{NW}\sin t^\circ_{\rm lab}\sin  \lambda_{\rm lab} +  \sin\theta_{NW}\cos t^\circ_{\rm lab} \Big\} ~~ + 0.4602\cos\theta_{NW}\cos  \lambda_{\rm lab}\nonumber \\
&& \label{eq:galacticZ}
\end{eqnarray}

Let us consider $\hat{\vec{X}}_{\rm{DP}} = \Big(\sin\theta\cos\phi, \,\sin\theta\sin\phi, \,\cos\theta\Big)$ a unit vector of the DP polarization in the galactic frame. The angle $\alpha$ between the DP polarization and the tangent of mirror is the complement of the angle with respect to $\hat{\vec{Z}}$:
\begin{equation}
	\cos^2\alpha = 1 - \Big(\hat{\vec{X}}_{\rm{DP}}\cdot \hat{\vec{Z}}\Big)^2
	\label{eq:cos2}
\end{equation} 

\bibliographystyle{JHEP}
\bibliography{references}

\end{document}